\documentclass[11pt]{article} \textwidth 165mm \textheight 240mm
\usepackage{amsfonts}

\font\mybb=msbm10 at 12pt
\def\bb#1{\hbox{\mybb#1}}

\begin{document}
\thispagestyle{empty}

\begin{flushright}
arXiv:0907.4681[hep-th] \\
July 27, 2009. \\ V2: August 11, 2009
\end{flushright}

\vskip 5.5cm

\begin{center}
\baselineskip=16pt {\LARGE\bf On superembedding approach to multiple
D-brane system. D0 story. }

\vskip 1.5cm

\renewcommand{\thefootnote}{\fnsymbol{footnote}}

 {\large\bf Igor A. Bandos$^\dagger{}$}\footnote{
{ Also at  A.I. Akhiezer Institute for Theoretical Physics,
 NSC Kharkov Institute of Physics $\&$ Technology,
UA 61108,  Kharkov, Ukraine. }  {E-mail: igor\_bandos@ehu.es,
bandos@ific.uv.es}} \vskip 1.5cm {\it $^\dagger$ Department of
Theoretical Physics, University of the Basque Country,  \\ P.O. Box
644, 48080 Bilbao, Spain
\\ and IKERBASQUE, the Basque Foundation for Science, 48011, Bilbao, Spain}

\vspace{1.5cm}

\end{center}

\vskip 0.5cm

\begin{quote}

{ We develop superembedding approach to multiple D-particle
(D$0$-brane) system. In flat target $D=10$ type IIA superspace this
produces the supersymmetric and Lorentz covariant version of the
Matrix model equations. The equations following from our
superembedding approach to multiple D$0$ in curved type IIA
superspace shows the Myers 'dielectric brane effect', {\it i.e.}
interaction with higher form gauge fields which do not interact with
a single D$0$-brane.}

\end{quote}
\renewcommand{\thefootnote}{\arabic{footnote}}
\setcounter{footnote}{1}

\vskip 3.5cm
 \setcounter{page}{1}
 \newpage
\section{Introduction}
\setcounter{equation}{0}

Although the first appearance of D-branes is dated by late 80th
\cite{Sagnotti,Horava,Polchinski89,Leigh:1989jq}, their  special
r\^ole in String/M-theory was widely appreciated in middle 90th when
it was found that Dp-brane carries RR (Ramond-Ramond)  charges {\it
i.e.} interact with the antisymmetric tensor gauge fields $C_{p+1}$,
$C_{p-1}$, $\ldots$, with  respect to which the fundamental strings
is neutral \cite{Polchinski95}. It was quickly understood that the
low energy dynamics of multiple Dp-brane system is described by the
maximal supersymmetric $d=p+1$ gauge theory with the gauge group
$U(N)$ in the case of N D-branes \cite{Witten:1995im}. Already this
limit was quite productive \cite{D0-1996} and, in particular,
allowed the conjecture of M(atrix) theory that the  Matrix model
\cite{Banks:1996vh} considered as a theory of multiple D0-brane
system, could provide a nonperturbative treatment of the M-theory.

The complete nonlinear supersymmetric action for a single D$p$-brane
was constructed in \cite{Townsend95} for $p=2$ and in \cite{Dpac}
for general $p$\footnote{To be more precise, the actions of
\cite{Townsend95,Dpac} are complete modulo higher derivative
corrections.}. It contains the nonlinear Dirac--Born--Infeld (DBI)
term \cite{Witten:1995im,Townsend95,Tseytlin:BI-DBI} and the
Wess--Zumino (WZ) term  describing the coupling to RR gauge fields
\cite{D-braneWZ}. Even before the general actions were constructed
in \cite{Dpac},  the superembedding approach \cite{bpstv} was
developed for the case of Dp-branes in \cite{hs96} where it was
shown that that the supersymmetric equations of motion can be
obtained in its frame starting from the basic superembedding
equation (see below)\footnote{The complete form of the D$p$-brane
equations of motion can be found in \cite{bst97} and
\cite{IB+DS06}.}. For the discussion in this letter it will not be
excessive to notice that similar story happened with M5-brane: its
equations of motion had been derived in \cite{hs2}, in the frame of
superembedding approach, before the covariant and supersymmetric
action was constructed in \cite{Bandos:1997ui}.

As far as the nonlinear action for multiple D-brane systems is
concerned, it was expected that this should be described by some
non-Abelian generalization of the DBI plus WZ  action.  Tseytlin
proposed to use the symmetric trace prescription to construct the
non-Abelian DBI  action for the case of purely bosonic spacetime
filling D-brane \cite{Tseytlin:DBInA,Tseytlin:BI-DBI}.

Although the search for a supersymmetric generalization of such
non-Abelian DBI action has  not been successful, in 1999 Myers used
it as a starting point and applying a chain of dualities, derived
the so-called 'dielectric brane action' \cite{Myers:1999ps} which is
widely  accepted for the description of multiple D-brane system.
This action, however, does not possess neither supersymmetry nor
Lorentz symmetry. In spite of a number of attempts, its Lorentz
covariant and/or supersymmetric generalizations is not known in
general, although some progress was reached for the cases of low
dimensions $D$, low dimensional branes and low co-dimensional branes
\cite{Dima01,Dima+Panda,Drummond:2002kg}\footnote{Notice also
 very interesting {\it minus one quantization}
approach using {\it string with boundary fermions} proposed in
\cite{Howe+Linstrom+Linus}. The name 'minus one quantization' is
suggested by that, to reproduce Myers action in this scheme,  one
has to perform the  quantization of these boundary fermion sector.
We comment more on this approach in the concluding section.}.

As far as the superembedding approach shown its efficiency in
derivation of Dp-brane and M5-brane equations, it looks
natural to apply it in the search for equations of motion for the
multiple D$p$-brane system. In this letter we describe the results which this procedure gives  for the simplest case of multiple D$0$-brane system.

We begin by a very brief review of superembedding approach to single
D$p$-branes for arbitrary $p$, with particular emphasis on
D$0$-brane case, which provides a technical basis for our study.
Then we argue in favor of the  idea to search for the description of
multiple D$p$-brane systems by trying to define a possible nonlinear
generalization of the non-Abelian SYM multiplet by some set of
constraints on the D$p$-brane worldvolume superspace ${\cal
W}^{(p+1|16)}$. The embedding of this  worldvolume superspace with
$d=p+1$ bosonic and $16$ fermionic directions into the type II
target superspace $\Sigma^{(10|32)}$ is determined by the so-called
superembedding equation and, in the case of higher branes ($p\geq 6$
\cite{HS+Chu=PLB98}), also by constraints on the Abelian SYM gauge
field characteristic of D$p$-brane. We argue that the use of these
constraints is useful also for a lower $p$ cases.

Then, turning to the case of multiple D0-brane, we propose the d=1
${\cal N}=16$ SYM constraints which express its field strength in
terms of a nanoplet of $su(N)$ valued superfields ${\bb X}^i$
obeying a superembedding--like equation $D_\alpha {\bb X}^i=4i
(\sigma^{0i}\Psi)_\alpha$. The leading component of this superfield,
appearing in the expression for the dimension 1 (spinor-spinor)
field strength of the $SU(N)$ gauge superforms,
$G_{\alpha\beta}=\sigma^i_{\alpha\beta}{\bb X}^i$, describe the
relative motion of N D$0$-brane constituents of the system. We show
that they produce a nonlinear equations of motion, which, in the
case of flat target superspace, describe a non-Abelian D=10 SYM
dimensionally reduced to $d=1$ (the system which is used to define
the Matrix model). However, the superembedding approach is also able
to produce multiple D0-brane equations in an arbitrary type IIA
superspace supergravity background (and it is not clear how to
reproduce these equations just by SYM dimensional reduction). We
analyze the general algebraic structure of the bosonic equations of
motion for the multiple D0-brane in general type IIA supergravity
background, as follows from superembedding approach, and show that
these describe the Myers 'dielectric brane' effect of polarization
of multiple D$p$-brane system by external higher form fluxes,  {\it
i.e.} shows the  coupling of multiple D0-brane system to the higher
form gauge fields, which do not interact with a single D$0$-brane.
We conclude by discussion on our results and interesting directions
for future study.

{\bf 1.1. Basic notation.} Our notation are close to one in
\cite{IB08:Q7}. We denote the type II superspace coordinates by
$Z^{{\underline{M}}}= (x^\mu\, ,
\theta^{\check{\underline{\alpha}}})$ ($\mu=0,1,\ldots, 9$,
${\check{\underline{\alpha}}}=1,\ldots , 32$)  and supervielbein
form by
\begin{eqnarray}\label{Eua-cE}
{E}^{\underline{A}}:=
dZ^{\underline{M}}E_{\underline{M}}{}^{\underline{A}}(Z)=
 ({E}^{\underline{a}}, {\cal E}^{\underline{\alpha}})\; , \qquad {\cal E}^{\underline{\alpha}}=\cases{  (E^{\alpha 1}\, , \;
  E_{\alpha}^{2} \; ) \; for\; type \; IIA \cr
 (E^{\alpha 1}\, , \,
 E^{\alpha 2}) \; for\; type \; IIB } \;   \qquad
\end{eqnarray}
(${\underline{a}}=0,1,\ldots, 9$, ${\underline{\alpha}}=1,\ldots ,
32$). These are restricted by the set of supergravity constraints
the most essential of which are collected in the expression for the
bosonic torsion two form. For type IIA case these are
\begin{eqnarray}
\label{Ta=IIA} & T^{\underline{a}}:= DE^{\underline{a}} =
-i({E}^{1}\wedge \sigma^{\underline{a}} {E}^{1}+  {E}^{2}\wedge
\tilde{\sigma}^{\underline{a}} {E}^{2})\; , \qquad
\end{eqnarray}
while the constraints for type IIB superspace is obtained from
(\ref{Ta=IIA}) by omiting tilde (replacing
$\tilde{\sigma}^{\underline{a}}$ by $\tilde{\sigma}^{\underline{a}}$
in the second term in the brackets, $ T^{\underline{a}}:=
DE^{\underline{a}} = -i ({E}^{1}\wedge \sigma^{\underline{a}}
{E}^{1} + {E}^{2}\wedge \sigma^{\underline{a}} {E}^{2})$. Here
$\sigma^{\underline{a}}:=
\sigma^{\underline{a}}_{\alpha\beta}=\sigma^{\underline{a}}_{\beta\alpha}$
and $\tilde{\sigma}_{\underline{a}}:=
\tilde{\sigma}_{\underline{a}}^{\alpha\beta}=\tilde{\sigma}_{\underline{a}}^{\beta\alpha}$,
are $D=10$ Pauli matrices which obey $\sigma^{\underline{a}}
\tilde{\sigma}^{\underline{b}}+\sigma^{\underline{b}}\tilde{\sigma}^{\underline{a}}=2\eta^{\underline{a}\underline{b}}
$.

\section{Superembedding approach to a D$p$-brane.  }
\setcounter{equation}{0}

\subsection{Worldvolume superspace $W^{(p+1|16)}$}

Following the so--called doubly supersymmetric twistor-like approach
to superparticles and superstrings
\cite{stv,N=2ssp,DGHS93}\footnote{ See \cite{Dima99} for the review
and more references.} the superembedding approach
\cite{bpstv,hs96,hs2,bst97,Dima99,IB08:Q7} describes the dynamics of
super-p-brane in terms of embedding of a {\it worldvolume
superspace} into the {\it target superspace}. In the case of D=10
D$p$--branes (Dirichlet super-$p$-branes) the worldvolume superspace
${\cal W}^{(p+1|16)}$ has $d=p+1$ bosonic and $16$ fermionic
dimensions. We denote the local coordinates of ${\cal W}^{(p+1|16)}$
by $\zeta^{{\cal M}}=(\xi^m,\eta^{\check{\alpha}})$ ($m=0,1,.., p$,
${\check{\alpha}}=1,...,16$) the embedding of ${\cal W}^{(p+1|16)}$
into the $D=10$ type II target superspace $\Sigma^{(10|32)}$ can be
described in terms of coordinate functions
$\hat{Z}^{{\underline{M}}}(\zeta)=
(\hat{x}{}^{\underline{m}}(\zeta)\, ,
\hat{\theta}^{\check{\underline{\alpha}}}(\zeta))$,
\begin{eqnarray}
\label{WinS}  W^{(p+1|16)}\in \Sigma^{(10|32)} &:& \qquad
Z^{\underline{M}}= \hat{Z}^{\underline{M}}(\zeta) = (\hat{x}^{\underline{m}}(\zeta)\;
, \hat{\theta}^{\check{\underline{\alpha}}}(\zeta ))\; . \qquad
\end{eqnarray}

\subsection{The superembedding equation}

A particular beauty of the superembedding approach is that, for all
known super-$p$-branes, the embedding of the worldvolume superspace
into the target superspace is characterized by a universal equation
which is called the {\it superembedding equation}. This geometrical
equation (the name 'geometrodynamic equation' was used in
\cite{DGHS93}) restricts the coordinate functions
$\hat{Z}^{\underline{M}}(\zeta)$ and, in some cases, completely
determines the dynamics of superbrane.

To write the most general form of this superembedding equation let
us denote the supervielbein of $W^{(p+1|16)}$ by
\begin{eqnarray}
\label{eA=ea+} e^A= d\zeta^{{\cal M}} e_{{\cal M}}{}^{A}(\zeta) =
(e^a\; , \; e^\alpha) \; , \qquad a=0,1,\ldots , p\; , \qquad
\alpha=1,\ldots, 16 \; ,
\end{eqnarray}
and write the general decomposition of the pull--back of the
supervielbein of target type II superspace,
$\hat{E}^{\underline{A}}:=E^{\underline{A}}(\hat{Z})$ on this basis,
\begin{eqnarray}
\label{hEa=b+f}
 \hat{E}^{\underline{A}}:= E^{\underline{A}}(\hat{Z})=
d\hat{Z}^{\underline{M}}
E_{\underline{M}}{}^{\underline{A}}(\hat{Z}) = e^b \hat{E}_b^{\,
\underline{A}} + e^\alpha \hat{E}_{\alpha}{}^{\underline{A}} \; .
\qquad
\end{eqnarray}
The superembedding equation states that the bosonic supervielbein
form has zero projection on the worldvolume fermionic supervielbein
form. This is to say, it reads
\begin{eqnarray}
\label{SembEq}
 \fbox{$\hat{E}_{\alpha}{}^{\underline{a}}:= \nabla_{\alpha}
 \hat{Z}^{\underline{M}}\,
E_{\underline{M}}{}^{\underline{a}}(\hat{Z}) =0 \;$}\;  ,  \qquad
\nabla_\alpha:=e_\alpha^{{\cal M}}(\zeta) \partial_{{\cal M}}\, ,
\quad \zeta^{{\cal M}}=(\xi^m  ,  \eta^{\check{\alpha}} )\; .
\end{eqnarray}
It can be also presented in an equivalent form of
\begin{eqnarray}
\label{Ei=0} \hat{E}^{i}:= \hat{E}^{\underline{a}}
u_{\underline{a}}{}^i =0\; , \qquad
\end{eqnarray}
where $u_{\underline{a}} ^{\; i}=u_{\underline{a}} ^{\; i}(\zeta) $
are $(9-p)$ spacelike, mutually orthogonal and normalized
$10$-vector fields,
\begin{eqnarray}\label{uiuj=1-}
u_{\underline{a}} ^{\; i}u^{\underline{a}\; j}= -\delta^{ij}\; .
\qquad
\end{eqnarray}
Eq. (\ref{Ei=0}) means that they are orthogonal to the worldvolume
superspace. We can complete thier set till {\it moving frame} by
adding $d=(p+1)$ mutually orthogonal and normalized $D$-vector
fields $u_{\underline{a}} ^{\; b}=u_{\underline{a}}^{\; b}(\zeta)$
which are  tangential to the worldvolume superspace,
\begin{eqnarray}\label{utut=1tt}
u_{\underline{a}}^{\; a} u^{\underline{a}\; i}=0\; , \qquad  u_{\underline{a}} ^{\; a}
\eta^{\underline{a}\underline{b}}u_{\underline{b}} ^{\; b}=\eta^{ab}
\;  , \qquad a,b=0,1,\ldots , p\; , \qquad \underline{a}\, , \,
\underline{b}=0,1,\ldots , 9\; . \qquad
\end{eqnarray}
The statement that  $u_{\underline{a}}{}^{b}\;$ vectors are
tangential to the worldvolume superspace implies that their
contraction with the pull--back $\hat{E}^{\underline{a}}$ of the
target superspace bosonic supervielbein ${E}^{\underline{a}}$
provides us with a set of $d=(p+1)$ linearly independent
nonvanishing one-forms, which can be used as bosonic supervielbein
of the worldvolume superspace,
\begin{eqnarray}\label{Eua=ea}
\hat{E}^a:= \hat{E}^{\underline{b}} u_{\underline{b}} ^{\; a} = e^a
\; .  \qquad
\end{eqnarray}
This $e^a$ is referred to as induced by the (super)embedding. Eqs.
(\ref{Eua=ea}) and (\ref{Ei=0}) implies
\begin{eqnarray}\label{Eua=eua}
\hat{E}^{\underline{a}} = e^b u_b^{\; \underline{\,a}}
\; .  \qquad
\end{eqnarray}
This is one more equivalent form of the superembedding equation.

The fermionic supervielbein form $e^\alpha$ of the worldvolume
superspace $W^{(p+1|16)}$ can also be induced by superembedding.
When describing Dp-branes, it is convenient to identify $e^\alpha$
with the pull--back to $W^{(p+1|16)}$ of, say, the first of two
target space fermionic supervielbein forms
\begin{eqnarray}\label{Dp:ef=Ef1}
e^\alpha = \hat{E}^{\alpha 1}\; . \quad
\end{eqnarray}
Then the general decomposition of the second fermionic supervielbein
form reads
\begin{eqnarray}\label{Dp:Ef2=}
\cases{
 \hat{E}_{\alpha}^{2}=  e^\beta h_{\beta \alpha} + e^a \chi_{a\alpha} \qquad for \; IIA\; case   \cr \hat{E}^{\alpha 2}=  e^\beta h_\beta{}^\alpha + e^a \chi_a^\alpha \qquad for \; IIB\; case } \; . \quad
\end{eqnarray}

As far as the induced spin connection and normal bundle connections
are concerned,   it is convenient to write the definition of the
connections using the $SO(1,9)\times SO(1,p)\times SO(9-p)$
covariant derivative action on the moving frame vector,
\cite{bpstv},
\begin{eqnarray}\label{Dp:Du=Om}
Du_{\underline{b}}{}^a = u_{\underline{b}}{}^i \Omega^{ai}\; ,
\qquad Du_{\underline{b}}{}^i = u_{\underline{b}a} \Omega^{ai}\; .
\qquad
\end{eqnarray}
Both equations in  (\ref{Dp:Du=Om}) involve the covariant 1--form  $\Omega^{ai}$ which describes extrinsic geometry of $W^{(p+1|16)}$ embedded into the type IIB superspace and provides the supersymmetric (and superform) generalization of the so-called second fundamental form of the classical surface theory (see \cite{bpstv} for more discussion).

The selfconsistency condition for the superembedding equation Eq. (\ref{Ei=0}) gives, in particular, an algebraic equation for the spin tensor $h$ in (\ref{Dp:Ef2=}). For type IIA it reads
\begin{eqnarray}\label{hsih=-si}
 \, h\tilde{\sigma}^{\underline{b}}h^T u_{\underline{b}}{}^i
= -
 \sigma^{\underline{b}} u_{\underline{b}}{}^i
\; . \qquad
\end{eqnarray}
while for type IIB it is given by
$h\sigma^{\underline{b}}h^T u_{\underline{b}}{}^i =
- \sigma^{\underline{b}} u_{\underline{b}}{}^i $ (again $\tilde{\sigma}^{\underline{b}}\mapsto {\sigma}^{\underline{b}}$ rule).

\subsection{Constraints for the worldvolume gauge field. }

The constraints for the worldvolume gauge field strength of the
$Dp$-brane have the universal form
\begin{eqnarray}\label{Dp:F2=dA-B2=}
F_2:= dA -\hat{B}_2 = {1\over 2} e^b\wedge e^a F_{ab} \; ,  \qquad
\end{eqnarray}
where $\hat{B}_2$ is the pull--back  to the worldvolume superspace
$W^{(p+1|16)}$  of the type IIB NS-NS superform potential $B_2$. The
field strength of this is restricted by the constraints which, for
type IIA case, can be collected in the following differential form
expression
\begin{eqnarray}
\label{H3=IIA} & H_{3}:=dB_2 = - i {E}^{\underline{a}}\wedge
({E}^{1}\wedge \sigma_{\underline{a}} {E}^{1} -  {E}^{2}\wedge
\tilde{\sigma}_{\underline{a}} {E}^{2}) + & {1\over 3!}
{E}^{\underline{c}_3}\wedge{E}^{\underline{c}_2}\wedge
{E}^{\underline{c}_1}
H_{\underline{c}_1\underline{c}_2\underline{c}_3}\; . \qquad
\end{eqnarray}

The lowest dimensional (dim 2, $\propto e^\gamma\wedge e^\beta
\wedge e^a$) component of the Bianchi identities
\begin{eqnarray}\label{Dp:dF2=-H3}
dF_2= -\hat{H}_3 \;
\end{eqnarray}
 implies
\begin{eqnarray}\label{hsah=ska}
\matrix{ \, h\sigma^{\underline{b}}h^T u_{\underline{b}}{}^a =
\sigma^{\underline{b}} u_{\underline{b}}{}^c k_c{}^a   \qquad for \; IIB \; , \qquad \cr
\, h\tilde{\sigma}^{\underline{b}}h^T u_{\underline{b}}{}^a =
 \sigma^{\underline{b}} u_{\underline{b}}{}^c k_c{}^a  \qquad for\; IIA \; ,\qquad}
 k_a{}^b:=(\eta +F)_{ac}(\eta-F)^{-1}{}^{cb} \; ,
 \; . \qquad
\end{eqnarray}
Notice that this equation relates the spin-tensor $h$, appearing in
the decomposition of the pull--back of fermionic vielbein, and the
gauge field strength $F_{ab}$. One can easily check that the matrix
$k$ constrcuted from $F_{ab}$ as in (\ref{hsah=ska}) is SO(1,p)
group valued, {\it i.e.} it obeys $k\eta k^T=\eta$
\cite{ABKZ,IB+DS06},
 \begin{eqnarray}\label{kinSOp}
 k =(\eta +F)(\eta-F)^{-1}\qquad \in \qquad  SO(1,9)\; . \qquad
\end{eqnarray}

Further study shows that the system of superembedding equation plus
the worldvolume gauge field constraints (\ref{Dp:F2=dA-B2=}) always
contain the dynamical equations among their consequences (and for
$p\leq 6$ D$p$-branes \cite{HS+Chu=PLB98} the superembedding
equation along suffice for this purposes). However, the details of
derivation are $p$-dependent so that we turn now to the case of
D$0$--brane which is of our main interest here. \footnote{This is
the place to note that the off-shell worldline superfield
formulations of massive $N=2$ superparticles in $D=2,3$ and $4$,
which are the lower-dimensional (and lower supersymmetric)
counterparts of a single D0--brane, were first considered in
\cite{N=2ssp}. }

\section{D$0$-brane in superembedding approach}

In the case of D$0$--brane there are nine spacelike directions
orthogonal to worldline and the tangent to the worldline gives a
time-like directions, so that the corresponding moving frame vectors
$(u_{\underline{a}}{}^0\, , u_{\underline{a}} ^{\; i})$ obey
\begin{eqnarray}\label{D0:uu=1}
u_{\underline{a}} ^{\; 0}u^{\underline{a}0}=1\; , \qquad
u_{\underline{a}} ^{\; i}u^{\underline{a}0}=0\; , \qquad
u_{\underline{a}} ^{\; i}u^{\underline{a}j}=-\delta^{ij}\; . \qquad
\end{eqnarray}
The worldvolume superspace $W^{(1|16)}$ has only one bosonic
direction, $e^a\mapsto e^0$ and the superembedding equation can be
written  as  (see (\ref{Eua=eua}))
\begin{eqnarray}\label{Eua=e0u0a}
\hat{E}^{\underline{a}} = e^0 u_0^{\, \underline{\,a}}
\; ,  \qquad
\end{eqnarray}
while the fermionic supervielbein forms read (see (\ref{Dp:ef=Ef1}))
\begin{eqnarray}\label{D0:ef=Ef1}
\hat{E}^{\alpha 1}=  e^{\alpha}\; , \qquad
 \hat{E}_{\alpha}^{2}=  e^\beta h_{\beta \alpha} + e^0\chi_{\alpha} \; . \quad
\end{eqnarray}
It is convenient to write  the selfconsistency conditions for the superembedding equation
(\ref{Eua=e0u0a}) in the form of
 \begin{eqnarray}\label{D0:hsih=}
 h{\tilde{\sigma}}{}^ih^T  =- {\sigma}^i
\; . \qquad
\end{eqnarray}
 using  the simplified notation
\begin{eqnarray}\label{D0:sui=si}  {\sigma}^0_{\alpha
\beta}:= {\sigma}^{\underline{b}}_{\alpha
\beta}u_{\underline{b}}{}^0 \; , \qquad {\sigma}^i_{\alpha
\beta}:= {\sigma}^{\underline{b}}_{\alpha
\beta}u_{\underline{b}}{}^i \; . \qquad
\end{eqnarray}
These are suggestive as far as the matrices (\ref{D0:sui=si}) and
$\tilde{\sigma}^0_{\alpha
\beta}:= \tilde{\sigma}{}^{\underline{b}}_{\alpha
\beta}u_{\underline{b}}{}^0 $, $\tilde{\sigma}^i_{\alpha
\beta}:= \tilde{\sigma}{}^{\underline{b}}_{\alpha
\beta}u_{\underline{b}}{}^i $ do possess the algebraic properties of D=10 Pauli matrices. However,
one should keep in mind that  they are not constant matrices but rather obey
\begin{eqnarray}\label{D0:Dsui=}
D{\sigma}_{\!_{\alpha \beta}}^0= {\sigma}_{\!_{\alpha
\beta}}^i\Omega^i\; , \qquad D{\sigma}_{\!_{\alpha \beta}}^i=
{\sigma}_{\!_{\alpha \beta}}^0\Omega^i\; , \qquad
\end{eqnarray}
where $\Omega^i$ is defined in (\ref{Dp:Du=Om}). In this notation the
the general solution of Eq.(\ref{D0:hsih=}) reads
\begin{eqnarray}\label{D0:h=} h_{\alpha \beta} = {\sigma}^{0}_{\alpha \beta}\; .
\qquad
\end{eqnarray}

This is the place to comment on  the worldvolume gauge field constraints for the
D$0$-brane  case (worldline gauge field).
For the $p=0$   the {\it r.h.s.} of Eq. (\ref{Dp:F2=dA-B2=}) clearly vanishes so
that the constraints read $F_2:=dA-\hat{B}_2=0$ and the Bianchi
identities (\ref{Dp:dF2=-H3}) simplify to $\hat{H}_3=0$. Their  only
nontrivial consequence reads
\begin{eqnarray}\label{D0:hsu0h=}
 h\tilde{\sigma}^{0}h^T  =
 {\sigma}^0\; .  \qquad
\end{eqnarray}
Eq. (\ref{D0:hsu0h=}) is satisfied identically by the general
solution (\ref{D0:h=}) of Eq. (\ref{D0:hsih=}). This shows  that the
gauge field constraints in the case of D$0$-brane are dependent,
which is in agreement with the known statement that the
superembedding equation alone is sufficient to describe dynamics in
this case. On the other hand, using both (\ref{D0:hsih=}) and
(\ref{D0:hsu0h=}), one finds the solution (\ref{D0:h=}) immediately,
much easier than using only (\ref{D0:hsih=}); this illustrates  that
the use of the superspace constraints for the worldvolume gauge
fields (\ref{Dp:F2=dA-B2=}) is helpful also in the cases when the
superembedding equation is sufficient to describe the brane
dynamics.

Another consequence of the selfconsistency conditions for the
superembedding equation (\ref{Eua=e0u0a}) is that $\Omega^i$ in
(\ref{D0:Dsui=}) is expressed by
\begin{eqnarray}\label{D0:Omi=}
\Omega^{i}= e^0\, K^i - 2i e^\beta
 ({\sigma}^0\tilde{\sigma}^i\chi)_\beta\;  \qquad
\end{eqnarray}
in terms of fermionic superfield
$\chi_\alpha=\hat{E}_0{}_\alpha^{2}$ and bosonic superfield
\begin{eqnarray}\label{D0:Ki:=}
 K^i&:= - u^i_{\underline{a}}D_0 \hat{E}_0^{\underline{a}}
\; , \qquad \hat{E}_0^{\underline{a}}:=
\nabla_0\hat{Z}^{\underline{M}}{E}_{\underline{M}}{}^{\underline{a}}(\hat{Z})
\; . \qquad
\end{eqnarray}
This latter has a meaning of mean curvatures of the D0-brane
(super)worldline in target type IIA superspace. The bosonic and
fermionic equations, which can be now obtained from the
selfconsistency condition for the fermionic conditions
(\ref{D0:ef=Ef1}), are formulated in terms of these superfields. In
flat target superspace the equations of motion imply vanishing of
both $\chi_\alpha$ and $K^i$,
\begin{eqnarray}\label{D0:flatEqm=} \chi_\alpha :=
\hat{E}_{_0}{}^2_{\alpha} =0 \; , \qquad
 K^i:= - u^i_{\underline{a}}D_{_0} \hat{E}_{_0}{}^{\underline{a}} =  0  \; .  \qquad
\end{eqnarray}
In general type IIA supergravity background the fermionic equations
of motion acquires the {\it r.h.s.}
\begin{eqnarray}\label{D0:DiracEq}
\chi_\alpha:= \hat{E}_{_0}{}^2_{\alpha} = \Lambda_\alpha  \;  \qquad
\end{eqnarray}
 defined by
\begin{eqnarray}\label{D0:L=L1+sL2}
 & \Lambda_\alpha:=  (\hat{\Lambda}_1-\hat{\Lambda}_2\sigma^0)_\alpha \;  ,
\qquad \Lambda_{\alpha 1} := {i\over 2}{(D_{\alpha 1}{\Phi})}\;  ,
\quad \Lambda_2^{\alpha}\; := {i\over 2}\,{(D_2^{\alpha}\,{\Phi})}\;
, \qquad
\end{eqnarray}
in terms of pull--backs of the Grassmann derivatives of the dilaton
superfield. The origin of this {\it r.h.s.} is nonvanishing
fermionic torsion of the target type IIA superspace
\cite{Bellucci+Gates+89}
\begin{eqnarray}\label{Tal1-2=IIA}
& T^{\alpha 1} =
 - 2i E^{\alpha 1}\wedge E^{\beta 1} \Lambda_{\beta 1} + i E^{1}\sigma^{\underline{a}}\wedge E^{1}\,
\tilde{\sigma}_{\underline{a}}^{\alpha\beta} \Lambda_{\beta 1} +
\propto E^{\underline{b}}  \, ,  \, \qquad \nonumber \\
 & T_{\alpha}^{ 2} =
 -   2i  E^2_{\alpha}\,\wedge
E^2_{\beta} \; \Lambda_2^{\beta}\,  + i
E^{2}\tilde{\sigma}_{\underline{a}}\wedge E^{2}\,
{\sigma}^{\underline{a}}_{\alpha\beta} \, \Lambda_2^{\beta } +
\propto E^{\underline{b}}   \; . \qquad
\end{eqnarray}
The bosonic equation for D$0$-brane in general supergravity
background reads
\begin{eqnarray}\label{D0:bEqm=}
 K^i&:= - u^i_{\underline{a}}D_0 \hat{E}_0^{\underline{a}} = {1\over 16}
 \tilde{\sigma}{}^{i\alpha\beta} (t_{\alpha\beta} - D_{\alpha} \Lambda_{\beta} )
 + {7i\over 8}(\hat{\Lambda}_2\sigma^{0i}\hat{\Lambda}_1) = \qquad \nonumber \\
 & =e^{\hat{\Phi}}\hat{R}{}^{0i}+\widehat{D^i\Phi}+ {\cal O}(fermi^2)\;  , \qquad
\end{eqnarray}
 where $t_{\alpha\beta}= \left( \widehat{T}_{\alpha 1\,
\underline{a}}{}^2_{\beta }+ \sigma^0_{ \alpha\gamma}
\widehat{T}^{\gamma}_{2}{}_{\underline{a}}{}^2_{\beta}-
\widehat{T}_{\alpha 1 \underline{a}}{}^{\delta 1} \sigma^0_{\delta
\beta }- \sigma^0_{ \alpha\gamma}
\widehat{T}^{\gamma}_{2}{}_{\underline{a}}{}^{\delta 1}
\sigma^0_{\delta \beta }\right)\, u^{\underline{a}\, 0}$. To arrive
at the second line of Eq. (\ref{D0:bEqm=}), written explicitly up to
the fermionic contributions, one has to use also the explicit
expressions for $D_{\alpha} \Lambda_{\beta}$ and dimension $1$
component of fermionic torsions in terms of the background fluxes;
these can be extracted from the results of \cite{Bellucci+Gates+89}
(see also \cite{IBatal}).

\section{Multiple D0-brane equations from superembedding approach.}

It is the usual expectation that the action for a system of N
D$p$-branes  will essentially be  a nonlinear generalization of the
U(N) SYM action. In particular, the (purely bosonic and not Lorentz
invariant) Myers action \cite{Myers:1999ps} is of this type. Then
the equations of motion which should follow from a hypothetical
supersymmetric and Lorentz covariant generalization (or
modification) of this action are expected to contain the SU(N) SYM
equations ($U(N)=SU(N)\times U(1)$) while the center of mass motion
is expected to  be described by a usual type of coordinate functions
$\hat{Z}^{\underline{M}}(\xi)$ and by related equations for the U(1)
gauge fields (presumably coupled to the SU(N) equations). Notice
that the center of mass equations of motion (and equations for U(1)
gauge fields which is expected to be involved in the center of mass
supermultiplet) are expected to be quite close to the equations  for
a single D$p$-brane, but with the single brane tension $T$ replaced
by $NT$. Our task now is to apply the superembedding approach in
search for such supersymmetric  equations in the simplest $p$=$0$
case.

\subsection{Non-Abelian
${\cal N}=16$, $d=1$ SYM constraints on D0-brane}
First, as far as
the superembedding description of one brane is based on
superembedding equation stating, in its form of (\ref{SembEq}), that
the pull--back of the target space bosonic vielbein to the
worldvolume superspace ${\cal W}^{(1|16)}$ do not have projections
on the fermionic vielbein of ${\cal W}^{(1|16)}$, it is natural to
expect that the center of mass motion of the system of multiple
D0-brane will also obey the superembedding equations \footnote{Of
course, this is not a proof. But the universality of the
superembedding equation, which is valid for all extended objects
studied till now in their maximal worldvolume superspace
formulations, and the difficulties one arrives at in any attempt to
modify it suggest to believe in its necessity. Finally, if a
modification of superembedding equation more appropriate do describe
multibrane systems were found, we hope that our present study would
be useful in search for such a hypothetical modification. }.

As far as the superembedding equation puts the D0--brane model on
the mass shell, our superembedding approach to ND$p$-brane model
predicts that the center of mass motion will be described by the
equations of motion of single brane with tension  $N\cdot T$.  (In
the case of D$0$-branes, {\it i.e.}  D-particles, $T$ has a meaning
of the particle mass). Then, in the light of the above discussion,
the only possibility to describe the multiple D0-brane system in the
framework of superembedding approach seems to be to consider a {\it
non-Abelian SU(N) gauge field supermultiplet} on the D0-brane
worldvolume superspace $W^{(1|16)}$. (See \cite{IB08:Q7} for more
discussion in the context of searching for hypothetical Q7-branes
\cite{Dima+Eric=2007}.)

 This can be defined by an $su(n)$ valued non-Abelian gauge potential
one form $A=e^0A_0 + e^\alpha A_\alpha$ with the field strength
\begin{eqnarray}\label{D0:G=dA-=}
G_2= dA - A\wedge A= {1\over 2}e^\alpha\wedge e^\beta
G_{\alpha\beta} + e^0\wedge e^\beta G_{\beta 0}\;  \qquad
\end{eqnarray}
To get a nontrivial consequences for the structure of the field
strengths $G_{\alpha\beta}$, $G_{\beta 0}$ one has to impose
constraints. A natural possibility  is
\begin{eqnarray}\label{D0:G=sX}
G_{\alpha\beta}= i \sigma^i_{\alpha\beta} {\bb X}^i \; , \qquad
\end{eqnarray}
with some $su(N)$ valued $SO(9)$ vector superfield ${\bb X}^i$. The
Bianchi identities $DG_2= dG_2 - G_2\wedge A + A \wedge G_2\equiv 0
$ are satisfied if  ${\bb X}^i$ obeys
\begin{eqnarray}\label{D0:DX=sPsi}
D_{\alpha} {\bb X}^i = 4i (\sigma^0\tilde{\sigma}{}^i)_{\alpha}{}^{
\beta} \, {\Psi}_\beta \; . \qquad
\end{eqnarray}
and $ G_{\alpha 0}= i\Psi_\alpha + {i\over 2}
(\sigma^{0i}\Lambda)_\alpha {\bb X}^i$. It is natural to call
(\ref{D0:DX=sPsi}) {\it superembedding like equation} as it gives a
non-Abelian $SU(N)$ generalization of the gauge fixed form of the
linearized superembedding equation (\ref{SembEq}) (this reads
$D_\alpha X^i= \propto
(\sigma^0\tilde{\sigma}^i(\Theta^2-\Theta^1))_\alpha$, see
\cite{hs96}).

\subsection{Multiple D0-brane equations of motion from $d=1$ ${\cal N}=16$ SYM constraints.
Flat target superspace. Relation to D=10 SYM and M(atrix) model. }
Let us, for simplicity, consider the case of flat target type IIA
superspace, in which, on the mass shell of D0-brane, $\Omega^i=0$,
so that $\sigma^0_{\alpha\beta}$ and $\sigma^i_{\alpha\beta}$ are
covariantly constants, $D\sigma^0_{\alpha\beta}=0
=D\sigma^i_{\alpha\beta}$. In this case the integrability conditions
for Eq. (\ref{D0:DX=sPsi}) ($D_{(\beta}D_{\alpha )}{\bb X}^i=...$)
 result in
\begin{eqnarray}\label{D0:DPsi=}
D_{\alpha}\Psi_{\beta } = & -{1\over 2} \sigma^i_{\alpha\beta}
D_0{\bb X}^i+{1\over 16} \sigma^{0ij}_{\alpha\beta}
 [{\bb X}^i\, , \, {\bb X}^j ]\;   \qquad
\end{eqnarray}
and the integrability conditions for Eq. (\ref{D0:DPsi=}), result in
1d Dirac equation of the form
\begin{eqnarray}\label{D0:suDirac}
& D_{0}\Psi_{\beta}+ {1\over 4} [ (\sigma^{0j}\Psi)_{\beta}\, , \,
{\bb X}^j ]=0 \; .  \qquad
\end{eqnarray}
Applying the Grassmann covariant derivative $D_\alpha$ to the
fermionic Eq. (\ref{D0:suDirac}), one derives, after some algebra,
the following set of equations
\begin{eqnarray}\label{D0:D0D0Xi=}
D_0D_0{\bb X}^i - {1\over 32} [[{\bb X}^i\, , \, {\bb X}^j ]\, , \,
{\bb X}^j ]+ {i\over 8} \{ \Psi_\alpha , \Psi_\beta
\} \,\tilde{\sigma}^{i \alpha\beta}=0 \; , \qquad \\
\label{D0:D0XiXi=} [D_0{\bb X}^i , {\bb X}^i ] - 4i \{ \Psi_\alpha ,
\Psi_\beta \} \,\tilde{\sigma}^{0\alpha\beta}=0 \; .  \qquad
\end{eqnarray}
Eq. (\ref{D0:D0D0Xi=}) is a candidate bosonic equation of motion of
multiple D0-brane system. Eq.  (\ref{D0:D0XiXi=}) has the meaning of
1d Gauss low; this appears in gauge theories as an equation of
motion for the time component of gauge potential (which usually
plays the r\^ole of Lagrange multiplier).

The appearance of the counterpart of Gauss low (\ref{D0:D0XiXi=}),
characteristic of gauge theory,   is not occasional. The point is
that our equations appear to be the $D=10$ SYM equations
dimensionally reduced to $d=1$. The reason beyod this is that  our
constraints (\ref{D0:G=sX}) for $d=1$, ${\cal N}=16$ SYM multiplet
on the {\it flat} $d=1$ ${\cal N}=16$ superspace (as appears to be
the  worldvolume superspace of D$0$-brane when embedded in flat type
IIA superspace) can be obtained as a result of dimensional reduction
of $D=10$ supersymmetric gauge theory. Indeed, the $D=10$ SYM
constraints imply vanishing of spinor-spinor component of the field
strength, ${\bb F}_{\alpha\beta}:= 2{\bb D}_{(\alpha} {\bb A}_{\beta
)} + \{ {\bb A}_{\alpha},  {\bb A}_{\beta }\}- 2i
\sigma^{\underline{a}}_{\alpha\beta}{\bb A}_{\underline{a}}=0$.
Assuming independence of fields on the nine spacial coordinate, one
finds that spacial components ${\bb A}_{i}$ of the ten-dimensional
field strength are covariant and can be treated as scalar fields
${\bb A}_{i}=1/2{\bb X}^i$. Then the minimal covariant field
strength for $d=1$ SYM can be defined as $G_{\alpha\beta}:= 2{\bb
D}_{(\alpha} {\bb A}_{\beta )} +  \{ {\bb A}_{\alpha},  {\bb
A}_{\beta }\} - 2i \sigma^{\underline{0}}_{\alpha\beta}{\bb
A}_{\underline{0}}$ and, due to the original D=10 SYM constraints,
this is equal to $i\sigma^i{\bb X}^i$, as in Eq. (\ref{D0:G=sX}).

The above observation is important, in particular, because it
indicates the relation with Matrix model \cite{Banks:1996vh}.
Indeed, this is  described by the Lagrangian obtained by
dimensional reduction of the $D=10$ SYM down to $d=1$
\cite{Banks:1996vh}. Actually, the $d=1$ dimensional reduction of
the $D=10$ SYM was the first model used to describe D0-brane
dynamics in \cite{D0-1996} even before the complete action for
super-Dp-branes where constructed in \cite{Dpac}.

To resume, for the multiple D0-brane system in flat target type IIA
superspace the worldvolume superspace ${\cal W}^{(1|16)}$ is flat
and our superembedding approach results in equations which are
equivalent to the ones obtained as a result of dimensional reduction
of D=10 SYM and coincide with the Matrix model equations. However,
it can also be used to describe the multiple D0-brane system  in
curved supergravity background, where the way through 10D SYM
dimensional reduction is obscure.

\section{Multiple D0-branes in curved type IIA background. Polarization by external fluxes. }

In the case of worldvolume superspace of D0--brane moving in curved
target type IIA superspace the calculations become more complex due
to the presence of bosonic and fermionic background superfields. For
instance, instead of (\ref{D0:DPsi=}),  one finds
\begin{eqnarray}\label{D0:DPsi=IIA}
& D_{\alpha}\Psi_{\beta}  =  -{1\over 2} \sigma^i_{\alpha\beta} + {1\over 16}
\sigma^{0ij}_{\alpha\beta}[{\bb X}^i\, , \, {\bb X}^j ] +
\hat{\Lambda}_{1\epsilon}\Psi_\delta \Sigma_{1}{}^{\epsilon\delta}{}_{\alpha\beta}
+ (\hat{\Lambda}_{2}\sigma^0)_{\epsilon}\Psi_\delta \Sigma_{2}{}^{\epsilon\delta}{}_{\alpha\beta}
\;   \qquad
\end{eqnarray}
with spin-tensors $\Sigma_{1,2}{}^{\epsilon\delta}{}_{\alpha\beta}$
possessing the properties
$\sigma^{\underline{a}\underline{b}}{}_{\delta}{}^{\alpha}
\Sigma_{1,2}{}^{\epsilon\delta}{}_{\alpha\beta}\propto
\sigma^{\underline{a}\underline{b}}{}_{\beta}{}^{\epsilon}$ and
$D_\gamma \Sigma_{1,2}{}^{\epsilon\delta}{}_{\alpha\beta} \propto
\Lambda$. We will not need an explicit form of these (we leave this
and other details for future publication \cite{IBatal}) as our main
interest here will be in the algebraic structure of the bosonic
equations of motion.\footnote{The fermionic equations of motion in
the presence of fluxes have the structure of
\begin{eqnarray}\label{D0:Dirac=LR}
& {7\over 8}\left( D_{0}\Psi- {1\over 4} [ {\bb X}^i\, , \,
(\sigma^{0i}\Psi)]\right)=
(e^{\hat{\Phi}}\hat{R}{}^{0i}+\widehat{D^i\Phi})(\sigma^{0i}\Psi)
-{1\over 64} \sigma^{0k}\left(  - {1\over 2!}
e^{\hat{\Phi}}\hat{R}_{\underline{b}\underline{c}}
\sigma^{\underline{b}\underline{c}}- {1\over
4!}e^{\hat{\Phi}}\hat{R}_{\underline{b}\underline{c}\underline{d}\underline{e}}
\sigma^{\underline{b}\underline{c}\underline{d}\underline{e}}\right)
\sigma^{0k}\Psi +
 \qquad \nonumber \\ & +{1\over 64} \hat{H}^{0ij} \sigma^{0k}
\sigma^{ij} \sigma^{0k}\Psi +  D_0{\bb
X}^i \left(a_1 \sigma^{0i}\hat{\Lambda}_{ 1}
+ a_2 \sigma^{i}\hat{\Lambda}_2\right) +{1\over 16} [{\bb
X}^i, {\bb X}^j] \left(b_1
\sigma^{ij} \hat{\Lambda}_{ 1} - b_2
\sigma^{0ij}\hat{\Lambda}_2\right)  +
{\cal O}(\hat{\Lambda}_{1,2}\cdot \hat{\Lambda}_{1,2}\cdot \Psi)
\qquad \nonumber
\end{eqnarray}
with some constants $a_{1,2}$ and $b_{1,2}$.} Up to the fermionic
bilinears  proportional to the fermionic background fields these
read
\begin{eqnarray}\label{D0:D0D0Xi=IIA}
 D_0D_0{\bb X}^i - {1\over 32} [[{\bb X}^i\, , \, {\bb X}^j ]\, ,
\, {\bb X}^j ]+ {i\over 8} \{ \Psi_\alpha , \Psi_\beta \}
\,\tilde{\sigma}^{i \alpha\beta}=  D_0{\bb X}^j \, {\bb F}^{j,i} +
{1\over 16}[{\bb X}^j\, , \, {\bb X}^k]\,   {\bb G}^{jk,i} + \qquad \\
\nonumber + {\cal O}(\hat{\Lambda}_{1,2}\cdot \Psi)+ {\cal
O}(\hat{\Lambda}_{1,2}\cdot \hat{\Lambda}_{1,2})\; ,  \qquad
\end{eqnarray}
The SO(9) tensors   ${\bb F}^{j,i}$ and $ {\bb G}^{jk,i} $ in the
{\it r.h.s.} of (\ref{D0:D0D0Xi=IIA}) are expressed in terms of
fluxes by
\begin{eqnarray}\label{D0:FFij=}
{\bb F}^{j,i} = q_0 \widehat{D_0\Phi}\delta^{ij} + p_{1}
\hat{R}^{ij}+ q_{2} \hat{H}{}^{0ij}\; . \qquad
\\ \label{D0:GGijk=}  {\bb G}^{jk,i}= p_0\delta^{i[j}\widehat{D^{k]}\Phi}+ q_1 \delta^{i[j}\hat{R}{}^{k]0}+ p_2 \hat{H}{}^{ijk}+
q_3 \hat{R}{}^{0ijk}\; , \qquad
\end{eqnarray}
where $q_{0,1,2,3}$ and $p_{0,1,2}$ are constant coefficients
characterizing couplings to dilaton as well as to electric and
magnetic fields strength of 1-form, 2-form, 3-form gauge fields.

Notice that the center of mass motion is factored out and is
described by the single D0-brane equations (\ref{D0:bEqm=}),
\begin{eqnarray}\label{D0:bEqmDDXi=}
K^i:= D_0D_0\hat{X}^i + ... =
e^{\hat{\Phi}}\hat{R}^{0i}+\widehat{D^i\Phi}+ {\cal O}(fermi^2)\; , \qquad
\end{eqnarray}
($\hat{X}^i:=
\hat{Z}^ME_M^{\underline{a}}(\hat{Z})u_{\underline{a}}{}^i=\hat{X}^{\underline{a}}
u_{\underline{a}}{}^i+ ... $). Comparing this with Eq.
(\ref{D0:D0D0Xi=IIA}) we see that the multiple D0-branes, as
described by this equation, acquire interaction with higher form
'electric' and 'magnetic' fields
$\hat{H}^{0ij}:=H_{\underline{a}\underline{b}\underline{c}}(\hat{Z})u^{\underline{a}0}u^{\underline{b}i}u^{\underline{c}j}$,
$H^{ijk}:=H_{\underline{a}\underline{b}\underline{c}}(\hat{Z})u^{\underline{a}i}u^{\underline{b}j}u^{\underline{c}k}$,
$\hat{R}^{0ijk}:=R_{\underline{a}\underline{b}\underline{c}\underline{d}}(\hat{Z})
u^{\underline{a}0}u^{\underline{b}i}u^{\underline{c}j}u^{\underline{d}k}$.
As one D$0$-brane does not interact with these background, one may
say that the multiple D0-brane system is 'polarized' by the external
fluxes such that the interaction with higher brane gauge fields is
induced, much in the same way as neutral dielectric is polarized
and, due to this polarization, interacts with electric field. This
is the famous 'dielectric brane' effect observed by Myers in his
purely bosonic nonlinear action \cite{Myers:1999ps} which, thus, is
observed also for the D$0$--brane equations which have been obtained
from the superembedding approach.

\section{Conclusions and discussion}

In this letter we have reported the results of application of
superembedding approach to the search for multiple D$0$--brane
equations. For the case of arbitrary (on-shell) type II supergravity
background the dynamical equations obtained from the superembedding
approach describe the coupling of multiple D$0$--branes to the
higher NS-NS and RR fluxes ($H^{0ij}$, $H^{ijk}$ and $R^{0ijk}$).
Thus our equations of motion imply the 'polarization' of multiple
D0-brane system under external higher form fluxes which makes them
behaving like dipoles of charges characteristic for higher
D$p$-branes. This is the content of the so-called 'dielectric brane
effect' \cite{Myers:1999ps} characteristic for the (purely bosonic)
Myers action \cite{Myers:1999ps}. Further study of these equations
and of possible restrictions which they might put on the embedding
of multiple D$0$ into a given supergravity background and on their
interaction is an interesting problem for future study.

In the case of flat tangent superspace, when the background fluxes
vanish, the d=1, {\cal N}=16 worldvolume superspace of D0--brane is
flat and the  dynamical equations for the relative motion of
D$0$-brane 'constituents', which follows from the superembedding
approach,  are those of the D=10 SU(N) SYM dimensionally reduced
down to $d=1$.  They, thus,  actually coincide with what had been
used for the  very low energy description of multiple D0--brane
system \cite{D0-1996} and with the Matrix model equations
\cite{Banks:1996vh}.

The purely bosonic limit of our equations is clearly simpler than
the equations of motion following from the Myers action
\cite{Myers:1999ps}. In this sense, the suggestion of the
superembedding approach is that this simpler equations, together
with the single D0--brane equation describing the center of mass
motion, actually give the 'complete' description of the multiple
D0-brane system. The advantage of this description is that it is
supersymmetric and also Lorentz and diffeomorphism covariant, while
the Myers proposal \cite{Myers:1999ps} possesses neither of these
symmetries expected for a system of coincident Dp-branes.
Furthermore, as we have already stressed, our superembedding
approach also provides the completely supersymmetric and covariant
description of the 'dielectric brane effect'. \footnote{The Myers
action was (and is) motivated by that it is derived from T-duality.
But let us stress that the starting point for the corresponding
chain of duality transformations is the purely--bosonic D=10
non-Abelian Born-Infeld action based on the symmetric trace
prescription \cite{Tseytlin:BI-DBI}, and that supersymmetric
generalization of these 10D symmetric trace BI action is not known,
and its existence can be doubted (see discussion in
\cite{IB08:Q7}).}

Another important problem for future study is to understand the
relation of our results to the model in \cite{Dima+Panda}. As far as
our arguments in support of superembedding equation describing
center of mass motion of the multiple D$0$-brane system cannot be
considered as a rigorous proof, if a modification of the
superembedding equation resulting in a more complicated interaction
of D0-constituents did exist, a deeper understanding of the above
interrelation might suggest the way to obtain it; in this
hypothetical case our present study would provide a basis for such a
hypothetical modification.\footnote{To give an idea of the problems
one meets on the way of searching for generalization of our approach
which might incorporate nonlinear interactions suggested in
\cite{Dima+Panda}, let us notice that, although the consideration of
\cite{Dima+Panda} uses a purely bosonic worldline, the
identification of $\kappa$--symmetry with worldline supersymmetry
\cite{stv} can be used to identify the corresponding superembedding
equation. This appears to be $\hat{E}^i= \hat{C}_1 M^i$, where $M^i$
is some worldvolume function ($\propto {\partial M\over\partial
p^i}$ in the notation of \cite{Dima+Panda}) and $\hat{C}_1$ is the
pull--back of the RR 1-superform of the type IIA supergravity
($\propto d\theta_1 , \theta_2 - \theta_1 d\theta_2$ in the case of
flat superspace considered in \cite{Dima+Panda}). The problem with
such a generalization of superembedding equation is that it is not
invariant under the gauge symmetry of the RR 1-form.}

Notice that a modified superembedding equation does appear in the
boundary fermion approach of \cite{Howe+Linstrom+Linus}. However,
this happens for the superembedding equation describing the
embedding of a worldsheet superspace ${\cal W}^{(p+1|16+2N)}$
enlarged, to describe the 'classical' counterpart of multiple
D$p$-brane system but not just a single D$p$-brane, by $2N$
additional fermionic directions, into the standard type II
superspace $\Sigma^{(10|16+16)}$, and the nonvanishing {\it r.h.s.}
in the counterpart of (\ref{Ei=0}), $\hat{E}^i= e^{\dot{\alpha}}
\chi_{{\dot{\alpha}}}{}^i$, happens to be proportional to the new
fermionic supervielbein forms $e^{\dot{\alpha}}$
(${\dot{\alpha}}=1,\ldots , 2N$) corresponding to the new boundary
fermion directions of ${\cal W}^{(p+1|16+2N)}$. This, hence, cannot
be used as a suggestion in our case, when the boundary fermions are
not used.

This is the place to make a more general comment on the  the
approach of \cite{Howe+Linstrom+Linus} and to stress that our
conclusions are not in contradiction with this work. In the second
of the articles \cite{Howe+Linstrom+Linus} a prescription was
formulated how to reproduce the Myers action from a specific
boundary fermion action. Basically it consists in i) fixing some
specific  gauge on an additional component of gauge potential
related to the boundary fermion directions and, then, ii) replacing
Poisson brackets by commutators and  boundary fermions by matrices.
However, in distinction to the original boundary fermion action, the
result of this prescription appears to be, besides purely bosonic,
also non-covariant with respect to diffeomorphisms and Lorentz
symmetry. A prescription of quantizing the boundary fermion in such
a way that to reproduce supersymmetric and Lorentz and
diffeomorphism covariant result is not known. In our opinion this
problem, noticed already in the original articles
\cite{Howe+Linstrom+Linus}, might be related with an attempt to
quantize only the boundary fermion sector leaving the center of mass
degrees of freedom classical. A complete quantization of the
dynamical system \cite{Howe+Linstrom+Linus}, which provides a fine
classical (or 'minus one quantized') description of the dynamical
system, should clearly result in an effective action describing,
besides D-branes, also supergravity degrees of freedom interacting
with them. The question whether it is possible to make a
quantization of a part of degrees of freedom and to arrive at a
covariant and supersymmetric description of multiple D-brane system
in this context is similar to the question of existence of multiple
D-brane action (without extra degrees of freedom), which was the
motivation of our present study.

\medskip

To conclude, as we have shown in this letter, the lowest
dimensional, $p=0$, multiple D$p$-brane system does allow for a
description in the frame of superembedding approach. It is
interesting to check whether such a description is possible for the
case of type IIB D$1$ (D-string) and type IIA D$2$-brane
(D-membrane). \footnote{More generally, it is interesting what is
the limiting value $p_{c}$ of the worldspace dimension $p$ above
which such a supersymmetric and Lorentz covariant description of
multiple of D$p$-brane dynamics is not possible. The study of
\cite{IB08:Q7} suggests that $p_c<7$, although a more direct check
would not be excessive.} We hope to turn to this problem in future
publications.

\section*{Acknowledgments}

The author is grateful to Paolo Pasti, Mario Tonin and especially
Dmitri Sorokin for useful discussions and comments,  as well as for
their kind hospitality at the Padova section of INFN and Padova
University at early stages of this work which was partially
supported by research grants from the Spanish MICINN (FIS2008-1980)
and the Ukrainian National Academy of Sciences and Russian
Federation RFFI (38/50--2008). The author also thanks Ulf
Lindstr\"{o}m and Warren Siegel for useful conversations and the
organizers of the 2009 Simons Workshop in Mathematics and Physics
for their hospitality at Stony Brook at the final stage of this
work.

 \end{document}